\documentstyle[editedvolume,psfig]{crckapb}

\def\msun{M$_\odot$}
\def\about{$\sim$}
\def\mdot{$\dot M$}
\def\approxlt{\ifmmode \rlap{$<$}{}_{{}_{{}_{\textstyle\sim}}} \else%
$\ifmmode \rlap{$<$}{}_{{}_{{}_{\textstyle\sim}}}$\fi}
\def\da{$\downarrow$}
\def\mc{\multicolumn}
\def\ni{\noindent}
\def\usec{$\mu$sec}
\def\hide#1{}
\begin{opening}
\title{\protect\vskip-30pt{\tt To appear in Proc. NATO ASI ``The
many faces neutron stars'', Lipari, Italy, 1996.}\protect\vskip20pt
KILOHERTZ QUASI-PERIODIC OSCILLATIONS \protect\\ 
IN LOW-MASS X-RAY BINARIES}

\author{M. VAN DER KLIS}
\institute{Astronomical Institute ``Anton Pannekoek'', \protect\\
University of Amsterdam}
\end{opening}

\runningtitle{KILOHERTZ QUASI-PERIODIC OSCILLATIONS}

\begin{document}

\section{Introduction}

The main motivation for studying X-ray binaries is not that they
exhibit a wide range of complex phenomenology, which they do, but that
they contain neutron stars and black holes. These are objects of
fundamental physical interest, as they make it possible to derive
information about the equation of state of very-high-density matter
and to perform tests of general relativity in the strong-field
regime. In this talk, I shall be discussing low-mass X-ray binaries
(LMXBs) containing neutron stars exclusively, as it is in the
understanding of the physics of these systems that great progress has
recently become possible by the discovery, with NASA's Rossi X-ray
Timing Explorer (RXTE), of a new phenomenon, kilohertz quasi-periodic
oscillations (kHz QPO).

In these LMXBs matter is transferred from a low-mass (\approxlt 1\,\msun) star to
a neutron star by way of an accretion disk. The X rays originate in the hot
(\about10$^7$\,K) plasma comprising the inner few 10$^1$ kilometers of the
flow. This is very close to the neutron star, which itself has a radius, $R$,
of order 10\,km, so that by studying the properties of this flow one expects
to be able to derive information about the star.

The high temperatures in the inner flow are caused by the release of large
amounts of gravitational energy when the matter descends into the neutron
star's very deep gravitational potential well ($GM/R\sim0.2c^2$; here and in
the remainder of this Section I assume a value of $1.4$\msun\ for the neutron
star's mass $M$). As the characteristic velocities near the star are of order
$(GM/R)^{1/2}\sim0.5c$, the dynamical time scale, the time scale for motion of
matter through the emitting region, is short; $\tau_{\rm dyn} \equiv
(r^3/GM)^{1/2}$$\sim$0.1\,ms for $r$=10\,km, and \about2\,ms for $r$=100\,km.

There is a number of processes that is expected to happen on these
time scales. Circular orbital motion takes place with a period
$P_{\rm orb}=2\pi\tau_{\rm dyn}$ (as seen from infinity; for a distant
observer Kepler's harmonic law holds in a Schwarzschild geometry). For
an orbital radius $r_{\rm orb}$ of 15\,km the orbital period is
$P_{\rm orb}$=0.8\,ms. If the star is small enough to allow it, orbital
motion at the general relativistic marginally stable orbit, at
$R_{6\!M}=6R_{\rm g} = 6GM/c^2$\about12.5\,km in a Schwarzschild geometry,
takes place at $P_{6\!M} = 2\pi6^{3/2}GM/c^3$ (as viewed from
infinity). This corresponds to a $P_{\rm orb}$ of \about0.6\,ms.
Including the general relativistic frame dragging due to the neutron
star spin this value decreases with the angular momentum of the
neutron star. Various oscillations of inner disk and star are expected
to take place on similar time scales, and of course the {\it spin} of
the neutron star is limited by this time scale as well (a realistic
neutron star maximum spin rate depends on gravitational radiation
losses for rapidly spinning neutron stars and is thought to be between
0.4 and 1.4\,ms; Cook, Shapiro \& Teukolsky 1994).

\begin{figure}[htb]
\begin{center}
\begin{tabular}{c}
\psfig{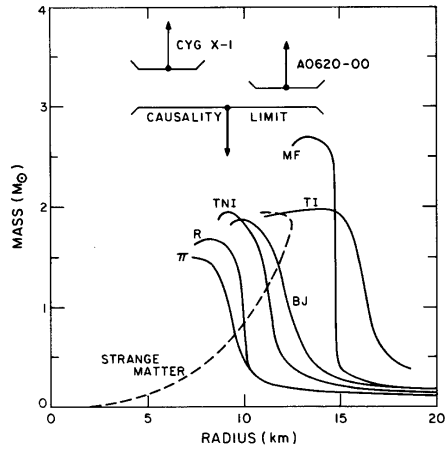}
\end{tabular}
\caption{\scriptsize Mass-radius diagram of neutron stars for various
different equations of state (McClintock 1986). \label{mrnsbh}}
\end{center}
\end{figure}

There are two different areas of fundamental physics to which study of
the accretion flows onto neutron stars can contribute. First, the
inner flow takes place in a region of spacetime where strong-field
general-relativistic effects are important. One may therefore expect
to be able to measure some of these effects, such as for example the
existence of a region where no stable orbits are possible (if the star
is small enough). Second, the interaction between the elementary
particles in the very high-density interior of neutron stars, which
determines the equation of state (EOS) of the matter there, is
insufficiently well understood to confidently predict the structure,
the maximum spin rate or the maximum mass of a neutron star, or even
the radius of a neutron star of given mass (Fig.\,\ref{mrnsbh}). This
means that, inversely, by measuring neutron star properties and
thereby constraining the EOS of very-high-density matter, one is
learning new things about the properties of elementary particles.

So, the main motivation for studying the flow of matter onto a neutron
star is that its nature depends on untested, or even unknown,
properties of spacetime and matter. Among the most basic properties of
these flows are their short, millisecond, characteristic time
scales. Up to less than a year ago, no direct information existed on
these flows' properties at these time scales. In this paper I report
on how, since February 1996, we are for the first time actually
observing time variability from these flows at the expected
millisecond time scales. New rapid-variability phenomena have been
discovered, namely quasi-periodic oscillations in the X-ray flux with
amplitudes of up to several 10\% of the total flux, quality factors
$Q\equiv\nu/\Delta\nu$ (see \S\ref{twinpeaks}) of up to nearly 1000, and frequencies
between \about300 and \about1200\,Hz. I shall call these phenomena
``kHz QPO'' (kilohertz quasi-periodic oscillations) throughout the
rest of this paper.

In studying this new phenomenon, the astrophysical community is building on
some 15 years of experience with studying similar, slower oscillation
phenomena in these same sources. I shall not attempt to provide a review here
of what happened in these years. For a description of time series power
spectral analysis methods I refer the interested reader to my paper (Van der
Klis 1989a) in the excellent ``Cesme'' proceedings book ``Timing Neutron
Stars'' in the same series as the present Volume. Have a look at this
paper if you want to find out if your QPO is significant and what its
fractional rms amplitude is.  The last pre-kHz-QPO overview of rapid X-ray
variability in X-ray binaries can be found in the book ``X-Ray
Binaries'' (Van der Klis 1995). Look here if you wish to find out about atoll
sources, Z sources and the latters' 16--60\,Hz horizontal-branch oscillations
and the 6--20\,Hz normal-flaring branch oscillations. For understanding what
follows, it is useful to remind the reader of the usual terminology with
respect to the subclasses of LMXBs (Hasinger \& Van der Klis 1989): Z sources
are near-Eddington accretors and probably have somewhat stronger (1--5
10$^9$\,G) magnetic fields, atoll sources are often X-ray burst sources, have
luminosities between 10$^{-3}\,L_{\rm Edd}$ and a few 10$^{-1}\,L_{\rm Edd}$, and
are thought to have somewhat weaker magnetic fields (10$^8$--10$^9$\,G). 

X-ray astronomers are presently scrambling to try and make sense of the
phenomenology of kHz QPO, which turn out to be at the same time highly
suggestive of interpretation and very restrictive of possible models, whereas
theorists have already begun working out sophisticated models. None of this
has reached an equilibrium state yet, and what I report in this paper will
necessarily be of a ``snapshot'' nature. What is clear at this point is that
for the first time we are seeing a rapid X-ray variability phenomenon that is
directly linked with a neutron star's most distinguishing characteristic (only
shared among macroscopic objects with stellar-mass black holes): its
compactness. This is particularly evident if the phenomena in some way are
related to orbital motion.  After all, a Keplerian orbital frequency 
$\nu_{\rm K} =
P_{\rm orb}^{-1}=(GM/4\pi^2r_{\rm K}^3)^{1/2}$ of 1200\,Hz 
around a 1.4\msun\ neutron
star as seen from infinity corresponds to an orbital radius $r_{\rm K} =
(GM/4\pi^2\nu_{\rm K}^2)^{1/3}$ of 15\,km, directly 
constraining the EOS and only
just outside the general-relativistic marginally stable orbit. Whatever the
model, for the first time we have to seriously worry about general
relativistic effects in describing the observable dynamics of the physical
system.

\section{Why now?}\label{why}

\begin{figure}[htb]
\begin{center}
\begin{tabular}{c}
\psfig{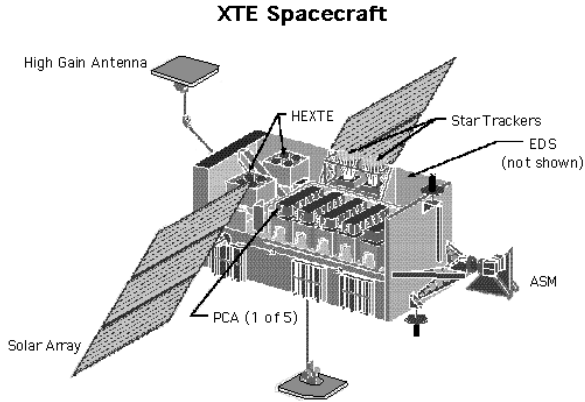}
\end{tabular}
\caption{\scriptsize The Rossi X-ray Timing Explorer (NASA). \label{xte}}
\end{center}
\end{figure}

It is appropriate to spend a few lines on the question: what made
these discoveries possible? On December 30th, 1995, NASA lauched the
Rossi X-ray Timing Explorer (RXTE). 
The proportional counter array (PCA)
onboard the RXTE satellite has an effective area of
\about7000\,cm$^2$. This is more than 4 to 8 times the area of the main
instrument of ESA's EXOSAT (1983--1986; half the detector array was
usually used to monitor the background) and nearly twice that of
ISAS's Ginga (1987--1991). For really bright sources, the PCA is able
to detect $>$10$^5$\,c/s, whereas the other two instruments could not
be safely pointed towards the brightest sources (these were only
observed off-axis) and attained highest count rates of
\about10$^4$\,c/s. The reason that this is such a crucial difference,
(but not the only reason that kHz QPO were not found before, see
below) is that the signal-to-noise ratio $n_\sigma$ (in sigmas) with which
photon-counting instruments detect weak variability of given
fractional rms amplitude $r$ goes {\it linear} with count rate $I_{\rm X}$:
for a power spectral feature of width $\Delta\nu$ and integration time
$T$ the signal-to-noise ratio is $n_\sigma =
{1\over2}I_{\rm x}r^2(T/\Delta\nu)^{1/2}$ (in the limit that
$T\gg\Delta\nu^{-1}$ and for negligible background) (Van der Klis
1989b; notice that this expression gives the signal-to-noise ratio, or
the ``single-trial detection-significance in sigmas'', which is not
the same as the sensitivity to QPO, which depends on the number of
trials; Van der Klis 1989a). If the background is not negligible, the
expression becomes $$n_\sigma = {1\over2}{S^2\over
B+S}r_S^2\left({T\over \Delta\nu}\right)^{1/2},$$ where $S$ and $B$ are
the source and background count rate, respectively, and $r_S$ is the rms
amplitude as a fraction of the {\it source} count rate. This means that for
a QPO peak to be detectable even at just a $>$3$\sigma$ level with
EXOSAT's ``half array'' it should be $>$25$\sigma$ in RXTE. Among the
kHz QPO peaks now known there are a few that are that significant,
but not many.

EXOSAT's best time resolution was 0.25\,ms, leading to a 2-kHz Nyquist
frequency, sufficient for all phenomena to be discussed here.  Ginga had 1\,ms
as its best time resolution (and even this was due to a last-minute design
change motivated by the discovery, with EXOSAT, of QPO phenomena in the
6--60\,Hz range), and therefore could not observe beyond 0.5\,kHz. Although
most of the new phenomena were therefore (just) out of its reach, with Ginga
the kHz QPO now known to occur in X-ray bursts in 4U\,1728$-$34, which have
frequencies of \about363\,Hz, could easily have been discovered. Yet EXOSAT
nor Ginga discovered the kHz QPO.

The reasons for this can be found in differences in telemetry
bandwidth and observing efficiency. Whereas sub-millisecond
time resolutions were sometimes employed in EXOSAT observations of 
the brightest sources (Z
sources and the brightest [``galactic center''] atoll sources, see
Hasinger \& Van der Klis 1989), they were rarely used for fainter
objects, and as it seems now, in the bright sources the kHz QPO have
lower fractional amplitudes and/or are restricted to the higher photon
energies. The reason that 0.25-ms resolutions were not routinely used
was that the limited telemetry bandwidth forced a choice between
instrumental modes stressing spectral resolution at the expense of
time resolution or {\it vice versa}. In order to get at least {\it
some} spectral resolution at millisecond time scales, observers often
favoured a mode (HER7) where 4-ms resolution data were obtained in 4
broad energy bands (plus a slow, high spectral-resolution mode running
in parallel). When searching for variability near 1\,ms in low-mass
X-ray binaries, they usually selected the brightest sources in the
class (the distinction between Z and atoll sources didn't become clear
until the EXOSAT mission was over; Hasinger \& Van der Klis
1989). See Jongert \& Van der Klis (1996) for an account of a
periodicity search in X-ray bursts observed with EXOSAT. The best
Nyquist frequencies in that search were indeed 2\,kHz, but in most
cases it was 0.25\,kHz or even less.

Most of the work with  Ginga concentrated on X-ray spectroscopy, not
timing. The high time-resolution modes could only be used when the
satellite was over the ground station in Japan, which made for a very
low observing efficiency in those modes. (This problem of
communication with the satellite is solved in the case of RXTE by
using the TDRSS communications satellites.) Consequently, relatively
short integration times $T$ were usually attained in high-time
resolution configurations. Nevertheless, Ginga data might yet yield a
kHz QPO as frequencies below its 0.5\,kHz Nyquist frequency do
occasionally occur (see Tables\,1 and 2).

\section{The early discoveries}

\begin{figure}[htbp]
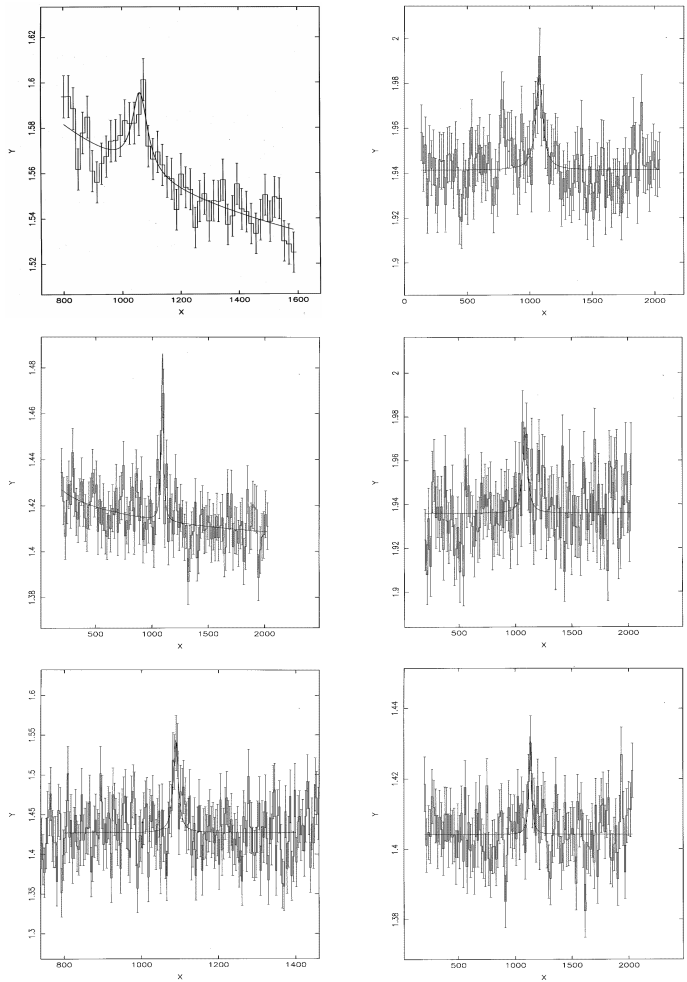

\begin{center}
\begin{tabular}{c}
\psfig{figure=obs1_16.postscript,width=4.2cm,height=4.2cm}
\hskip0.5cm
\psfig{figure=obs1_62.postscript,width=4.2cm,height=4.2cm}
\end{tabular}
\begin{tabular}{c}
\psfig{figure=obs2_16.postscript,width=4.2cm,height=4.2cm}
\hskip0.5cm
\psfig{figure=obs2_62.postscript,width=4.2cm,height=4.2cm}
\end{tabular}\begin{tabular}{c}
\psfig{figure=obs2_nb.postscript,width=4.2cm,height=4.2cm}
\hskip0.5cm
\psfig{figure=obs2_fb.postscript,width=4.2cm,height=4.2cm}
\end{tabular}
\caption{\scriptsize The first power spectra of Sco\,X-1 showing the
kHz QPO (Van der Klis et al. 1996a). Left to right, top to bottom,
respectively: first observation, 16 and 62\,\usec\ data; second
observation, 16 and 62\,\usec\ data; second observation, 250\,\usec\ 
data in NB and FB. Frequency in Hz along the X-axis; power (Leahy
normalized) along the Y-axis. An \about800\,Hz peak is visible in the
power spectra of the first observation, and was noticed at the time,
but was not sufficiently significant to report until the 16 and
62\,\usec\ data had been coherently combined (Van der Klis et
al. 1996c). \label{raw}}
\end{center}
\end{figure}

The first two sources in which kHz QPO were detected, during an
exciting few days in February 1996, were 4U\,1728$-$34 and
Sco\,X-1. Tod Strohmayer of Goddard Space Flight Center (GSFC) and
collaborators had proposed to observe 4U\,1728$-$34 in order to look
for neutron star r- and g-modes and for the neutron star spin. I had
traveled down to GSFC to perform the first RXTE observations of
Sco\,X-1, which our group (comprising some of the same people of the
RXTE PCA team as Tod's) had proposed to take advantage of RXTE's
10$^5$\,c/s capability (\S\ref{why}) which would improve the detection
significance of any weak variability by an order of magnitude. In our
proposal we had remarked that we suspected beat-frequency QPO
phenomena near ``\about700\,Hz'' might be observed. We, too, had
promised to look for the neutron star spin. While, after a bit of
waiting for our data to come out of the production pipeline, I was
just starting the first Sco\,X-1 analysis, Tod showed me his first
power spectrum of 4U\,1728$-$34 exhibiting a QPO peak near
800\,Hz. This was the first ``kHz'' phenomenon. When I got on with the
Sco\,X-1 data, sure enough, there was a peak near 1100\,Hz! This was
the first ``sub-millisecond'' phenomenon (Fig.\,\ref{raw}). 

Of course I discussed this result with Tod and the Sco\,X-1 team
members. We were very excited, but cautious, as obviously the new
effect might be instrumental, and our conversations focused on ways to
make certain it wasn't. I analyzed some of our data on Cyg\,X-1 which
had similar observational parameters and count rate as the
4U\,1728$-$34 data, and showed the result to the 4U\,1728$-$34 team:
it was clear that there was no sign of any 800\,Hz QPO there. For
Sco\,X-1 I found that the frequency of the 1100\,Hz QPO increased
along the so-called ``normal branch'' (NB) in the X-ray color-color
diagram, i.e., with mass accretion rate. Both teams were sufficiently
confident that the effects were real, IAU Circulars were submitted and
they came out on February\,20th (Van der Klis et al. 1996a; Strohmayer
et al. 1996a). In the Sco\,X-1 circular we said we thought we might be
seeing ``the keplerian frequency at the inner edge of the disk near
the magnetospheric boundary, or its beat frequency with a slower
(\about100\,Hz) pulsar''; in the 4U\,1728$-$34 circular Strohmayer et
al., more cautiously, remarked that ``explanations in terms of either
keplerian frequencies or a beat-frequency model cannot yet be ruled
out, although no evidence has yet been seen for a coherent pulsar
frequency in the same data.''. They also noted that ``oscillation
modes associated with the crust of a neutron star have frequencies in
the range observed''. Tod later told me that he submitted his
4U\,1728$-$34 circular before I submitted the Sco\,X-1 circular, and
that he was dismayed it came out second. Of course, officially the IAU
Central Bureau for Astronomical Telegrams is not in the business of
establishing scientific priorities; their mission is to rapidly
disseminate urgent information. Our discoveries were not independent,
as we discussed our results and assisted each other before
publication, and the two teams had common members. But let it be noted
for history: Tod saw the 800\,Hz in 4U\,1728$-$34 before I saw the
1100\,Hz in Sco\,X-1! I think it is good that in these exciting first
few days we focused on the science and freely exchanged information
rather than letting the competitive spirit get the better of us. As we
ventured on into the new field, this principle of a rapid and open
flow of information has been maintained, much helped by that
invaluable medium, the IAU Circulars. Particularly in the case of
satellite missions, with their limited life spans, the rapid
distribution of scientific results is essential to make optimal use of
the observational resources.

\begin{figure}[t]
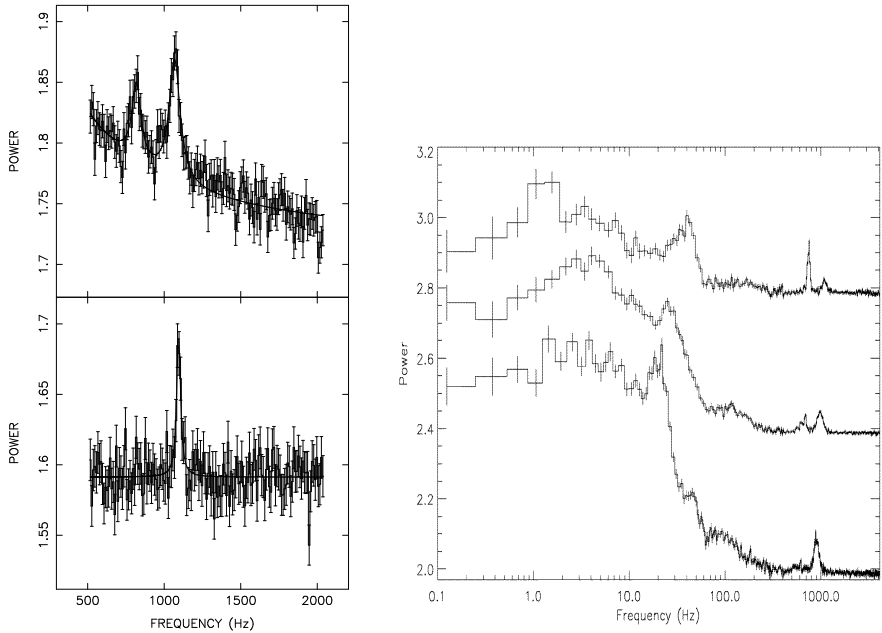

\begin{center}
\begin{tabular}{c}
\psfig{figure=apjl_xte_sco_fig2.postscript,width=4.5cm}
\hskip0.5cm
\psfig{figure=strohmayer_fig3.postscript,width=6.5cm,height=6.5cm}
\end{tabular}
\caption{\scriptsize Power spectra of Sco\,X-1 (left) 
showing double (top) and single (bottom) kHz QPO peaks (Van der Klis
et al. 1996c), and of 4U\,1728$-$34 (right), also with double and
single peaks in the kHz range (Strohmayer et al. 1996c). The sloping
continuum above 1\,kHz in Sco\,X-1 is instrumental. The peaks in the
1--50\,Hz range in 4U\,1728$-$34 are a known aspect of atoll source
phenomenology (Hasinger \& Van der Klis 1989). \label{discovery}}
\end{center}
\end{figure}

The next thing that happened was that on March\,6 Van Paradijs et
al. (1996) reported QPO in the 850--890\,Hz range in a third source,
4U\,1608$-$52. The first opportunity after all this excitement that
some of us had to come together and discuss the new discoveries was
during the 2-3 April RXTE Users Group Meeting at Goddard Space Flight
Center (Schwartz 1996). By then I had been able to detect a second QPO
peak near 830\,Hz in Sco\,X-1 simultaneous with the 1100\,Hz peak by
making use of a special capability of RXTE onboard data processing
(recording of ``double events''). In 4U\,1728$-$34 Todd Strohmayer
had an indication for a weak feature above 1000\,Hz, but he wasn't
sure of it yet. Just a few weeks later, he was: on April 26 in another
IAU Circular Strohmayer et al. (1996b) reported the simultaneous
presence in 4U\,1728$-$34 of two kHz QPO peaks whose frequencies
changed with time but whose difference remained near 363\,Hz. In
addition, they had seen a {\it third} QPO peak in five out of seven
X-ray bursts of the source which {\it also} had a frequency of 363\,Hz. Of
course, this showed that we were dealing with a beat-frequency
phenomenon (\S\ref{beat}). Strohmayer et al. (1996b) did not mention
this in so many words, but they did mention one of the most salient
conclusions of this: that the observations were consistent with a
neutron star spin period of 2.75\,ms (1/363\,Hz). Articles about these
initial Sco\,X-1 and 4U\,1728$-$34 results appeared in the September
20, 1996 ApJ Letters issue (Van der Klis et al. 1996c, Strohmayer et
al. 1996c; see Fig.\,\ref{discovery}).

\section{The current situation}\label{now}

\begin{table}[htbp]
\begin{center}\footnotesize\parskip=0pt\baselineskip=8pt
\renewcommand{\arraystretch}{0.7}\setlength{\doublerulesep}{1pt}
\caption{Observed frequencies of kilohertz QPO: atoll sources.} \label{atoll}
\bigskip
\begin{tabular}{lccccl}
\hline 
Source          & Lower  & Upper   & Peak       & Burst       & References\\
(in order       & peak   & peak    & sepa-      & QPO\\
of RA)          & freq.  & freq.   & ration     & freq.\\
                & (Hz)   & (Hz)    & (Hz)       & (Hz)\\ 
\hline
\noalign{\vskip0.2mm}
4U\,0614+09     &        & 520     &            &             & Ford et al. 1996, 1997a,b \\
                &        & \da     &            &             & Van der Klis et al. 1996d \\
                & 480    & 750     &            &             & M\'endez et al. 1997a \\
                & \da    & \da     & 327$\pm$4  & 328$^a$     & Vaughan et al. 1997 \\
                & 800    & 1150    &            &             & \\
\hline
\noalign{\vskip0.2mm}
4U\,1608$-$52   & \mc{2}{c}{\ 570} &            &             & Van Paradijs et al. 1996\\
                & \mc{2}{c}{\ \da }&            &             & Berger et al. 1996\\
                & \mc{2}{c}{\ 800} &            &             & Vaughan et al. 1997 \\
                & 650    & 940$^b$ & 293$\pm$7  &             & M\'endez et al. 1997b\\ 
                & \da    & \da     &            &             & Yu et al. 1997\\
                & 890    & 1125$^b$& 233$\pm$12 &             & \\
\hline
\noalign{\vskip0.2mm}
4U\,1636$-$53   & 830    &         &            &             & Zhang et al.  1996, 1997 \\ 
                & \da    &         &            &             & Zhang 1997 \\
                & 898    & 1147    & 257$\pm$20 &             & Van der Klis et al. 1996d \\ 
                & \da    & \da     &            & 581         & Wijnands et al. 1997a \\ 
                & 940    & 1190    & 276$\pm$10 &             & Vaughan et al. 1997 \\ 
                &        & \da     &            &             & \\
                &        & 1228    &            &             & \\      
\noalign{\vskip0.2mm}
                & \mc{2}{c}{\ 835} &            &             & \\
                & \mc{2}{c}{\ \da }&            &             & \\
                & \mc{2}{c}{\ 897} &            &             & \\
\hline
\noalign{\vskip0.2mm}
4U\,1728$-$34   &        &  500    &            &             & Strohmayer et al. 1996a,b,c \\
                &        & \da     &            &             & \\
                & 640    &  990    &            &             & \\
                & \da    & \da     & 355$\pm$5  & 363         & \\
                & 790    & 1100    &            &             & \\
\hline
\noalign{\vskip0.2mm}
KS\,1731$-$260  & 898    & 1159    & 260$\pm$10 & 524         & Morgan
\& Smith 1996 \\
                &        & \da     &            &             & Smith et al. 1997, Wijnands \\
                &        & 1207    &            &             & \& Van der Klis 1997 \\
\hline
\noalign{\vskip0.2mm}
4U\,1735$-$44     & \mc{2}{c}{\ 1140}&            &             & Wijnands et al. 1996, 1997b \\ 
                & \mc{2}{c}{\da   }&            &             & \\
                & \mc{2}{c}{\ 1160}&            &             & \\
\hline
\noalign{\vskip0.2mm}
MXB\,1743$-$29$^c$&        &         &            & 589         & Strohmayer et al. 1996d, \\
                &        &         &            &             & 1997 \\
\hline
\noalign{\vskip0.2mm}
4U\,1820$-$30   & 546    &         &            &             & Smale et al. 1996, 1997 \\
                & \da    &         &            &             & \\
                & 796    & 1065    & 275$\pm$8  &             & \\
\hline
\noalign{\vskip0.2mm}
Aql\,X-1        &\mc{2}{c}{\ 750}  &            &             & Zhang 1997 \\
                &\mc{2}{c}{\da}    &            & 549         &   \\
                &\mc{2}{c}{\ 830}  &            &             &   \\
\hline
\end{tabular}
\end{center}\vskip-1.2mm
Arrows indicate ranges over which the frequency was observed to vary;
these can \hfill\break\ni be made up of several overlapping ranges
from different observations. \hfill\break\ni Frequencies not connected
by arrows are measurements at different epochs.\hfill\break\ni
Frequencies in the same row were observed simultaneously (except for
burst QPO).\hfill\break\ni Entries straddling the upper and lower peak
columns are of single, unidentified peaks.\hfill\break\ni $^a$
Marginal; in persistent emission (Ford et al. 1997a). \hfill\break\ni
$^b$ Special detection method; M\'endez et al. (1997b).
\hfill\break\ni $^c$ Source identification uncertain.
\end{table}

Kilohertz QPO have now\footnote{October 1st, 1997} been reported in 13 LMXBs,
4 of which are Z sources and 9 of which are atoll sources and probable atoll
sources (see Van der Klis 1995 for a recent review of LMXB subclasses;
hereafter I shall use ``atoll source'' for LMXBs that probably fall in this
class as well as for those that definitely do so), together covering nearly
three orders of magnitude in X-ray luminosity (\about10$^{-3}$ to
\about1\,$L_{\rm Edd}$). Tables\,\ref{atoll} and \ref{z} summarize some of these 
results, and provide an overview of the literature that is
approximately complete as of this writing. A clear pattern of
systematic behaviour has emerged, and for that reason, rather than
getting into an exhaustive description of the phenomenology, or
following the historical line I shall concentrate on what I consider
at this point to be the main clues. For most of the kHz QPO
observational references in the remainder of this section I refer to
these Tables.

\subsection{Twin peaks}\label{twinpeaks}

\begin{figure}[htb]
\begin{center}
\begin{tabular}{c}
\psfig{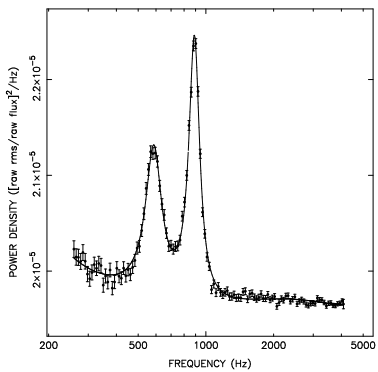}
\end{tabular}
\caption{\scriptsize Twin peaks in Sco\,X-1 (Van der Klis et al. 1997b). \label{scotwin}}
\end{center}
\end{figure}

\begin{figure}[htbp]
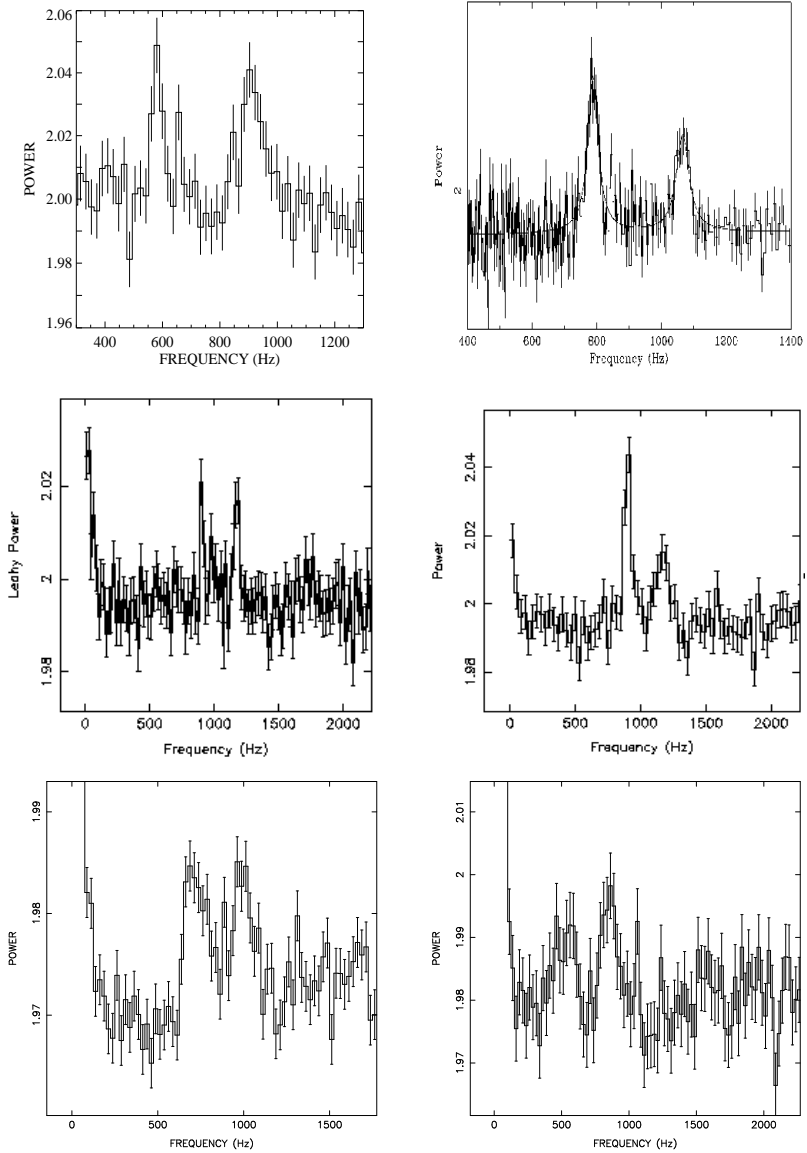

\begin{center}
\begin{tabular}{c}
\psfig{figure=0614_twinpeaks_ericford.postscript,width=5cm,height=5cm}
\hskip0.5cm
\psfig{figure=apjl_xte_1820m30_khz_discovery_smale_fig4.postscript,width=5cm,height=5cm}
\end{tabular}
\begin{tabular}{c}
\psfig{figure=apjl_ks1731_wijands_fig1_left.postscript,width=5cm,height=5cm}
\hskip0.5cm
\psfig{figure=apjl_1636_wijnands_fig2_left.postscript,width=5cm,height=5cm}
\end{tabular}
\begin{tabular}{c}
\psfig{figure=gx17p2_twinpeaks.postscript,width=5cm,height=5cm}
\hskip0.5cm
\psfig{figure=cygx2_twinpeaks.postscript,width=5cm,height=5cm}
\end{tabular}
\caption{\scriptsize Further examples of double kHz QPO peaks (``twin
peaks''). Top: the atoll sources 4U\,0614+09 and 4U\,1820$-$30 (Ford
et al. 1997a; Smale et al. 1997), middle: KS\,1731$-$260 and
4U\,1636$-$53 (Wijnands \& Van der Klis 1997; Wijnands et
al. 1997a) bottom: the Z sources GX\,17+2 and Cyg\,X-2 (Wijnands et
al. 1997c,d), respectively. \label{twin}}
\end{center}
\end{figure}

In nearly all kHz QPO sources (10 out of 13) {\it two} simultaneous
kHz peaks (hereafter: twin peaks) are observed in the power spectra of
the X-ray count rate variations
(Figs.\,\ref{discovery}--\ref{twin}). 

\begin{figure}[htbp]
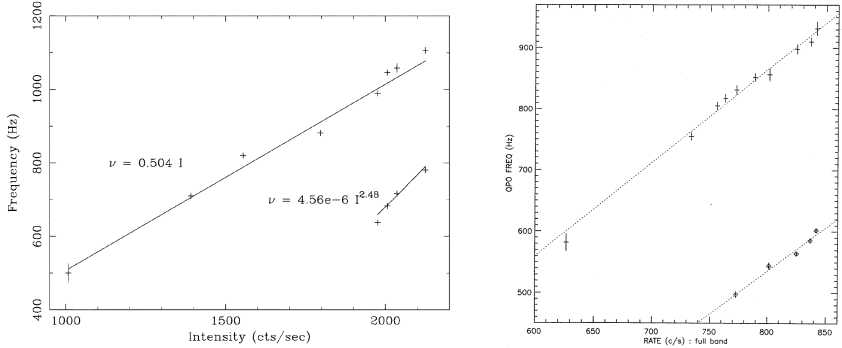

\begin{center}
\begin{tabular}{c}
\psfig{figure=apjl_strohmayer_1728_fig4.postscript,width=6cm}
\hskip0.5cm
\psfig{figure=0614_freq_vs_Ix.postscript,width=4.5cm}
\end{tabular}
\caption{\scriptsize Examples of frequency vs. count rate relations in
the atoll sources 4U\,1728$-$34 and 4U\,0614+09 (Strohmayer et
al. 1996c; Ford et al. 1996), respectively. In both cases the twin
peak separation was reported to be consistent with being
constant.\label{freq_Ix}}
\end{center}
\end{figure}

The lower-frequency peak
(hereafter the {\it lower peak}) has been observed at frequencies
between 325 and 940\,Hz, the higher-frequency peak (hereafter the {\it
upper peak}) has been observed at frequencies between 500 and
1228\,Hz. When the accretion rate \mdot\ increases, both peaks move to
higher frequency. In atoll sources \mdot\ is inferred to correlate
with X-ray count rate (except in some ``upper banana'' states,
Hasinger \& Van der Klis 1989, where no kHz QPO have so far been
seen), and in these sources kHz QPO frequency increases with count
rate (Fig.\,\ref{freq_Ix}). In Z sources in the so-called ``normal
branch'' (NB), \mdot\ is inferred to {\it anti}correlate to count
rate, and indeed in Z sources in the NB, the kHz QPO frequency increases
when the count rate drops, but in those sources, also, frequency
increases monotonically with
\mdot\ as inferred from curve length $S_Z$ along the Z track.

\begin{table}[htbp]
\begin{center}\footnotesize\parskip=0pt\baselineskip=8pt
\renewcommand{\arraystretch}{0.7}\setlength{\doublerulesep}{1pt}
\caption{Observed frequencies of kilohertz QPO: Z sources.} \label{z}\bigskip
\begin{tabular}{lcccl}
\hline 
Source & Lower & Upper & Peak & References\\ (in order & peak & peak &
sepa- \\ of RA) & freq.  & freq.  & ration \\ & (Hz) & (Hz) & (Hz) \\
\hline					       
\noalign{\vskip0.2mm}			       
Sco\,X-1        & 570    & 870     &           & Van der Klis et al. 1996a,b,c, \\
                & \da    & \da     & 292$\pm$2 & 1997b\\
                & 800    & 1050    & \da       & \\
                & \da    & \da     & 247$\pm$3 & \\
                & 830    & 1080    &           & \\
                &        & \da     &           & \\
                &        & 1130    &           & \\
\hline					       
\noalign{\vskip0.2mm}			       
GX\,5$-$1       &        & 567     &           & Van der Klis et al. 1996e \\
                &        & \da     &           & \\
                & 325    & 652     &           & \\
                & \da    & \da     & 327$\pm$11& \\
                & 448    & 746     &           & \\
                &        & \da     &           & \\
                &        & 895     &           & \\
\hline					       
\noalign{\vskip0.2mm}			       
GX\,17+2        &        & 645     &           & Van der Klis et al. 1997a \\
                &        & \da     &           & Wijnands et al. 1997c \\
                & 480    & 780     &           & \\
                & \da    & \da     & 294$\pm$8 & \\
                & 780    & 1080    &           & \\
                &        & \da     &           & \\
                &        & 1087    &           & \\
\hline
\noalign{\vskip0.2mm}
Cyg\,X-2        &        & 730     &           & Wijnands et al. 1997d \\
                &        & \da     &           & \\
                & 490    & 840     &           & \\
                & \da    & \da     & 343$\pm$21& \\
                & 530    & 860     &           & \\
                &        & \da     &           & \\
                &        &1020     &           & \\
\hline
\end{tabular}
\end{center}
Notes: see Table\,\ref{atoll}.
\end{table}

In the sources 4U\,0614+091 (Ford et al. 1997a; M\'endez et al. 1997a)
it has been observed that the frequency vs. count rate relation was
different at different epochs; the frequency seems to better track the
black-body flux as obtained from a two-component spectral fit than the
total count rate (Ford et al. 1997b). In 4U\,1608$-$52 (M\'endez et
al. 1997b) the frequency vs. count rate relation is either
non-monotonic or time-variable.

\begin{figure}[htbp]
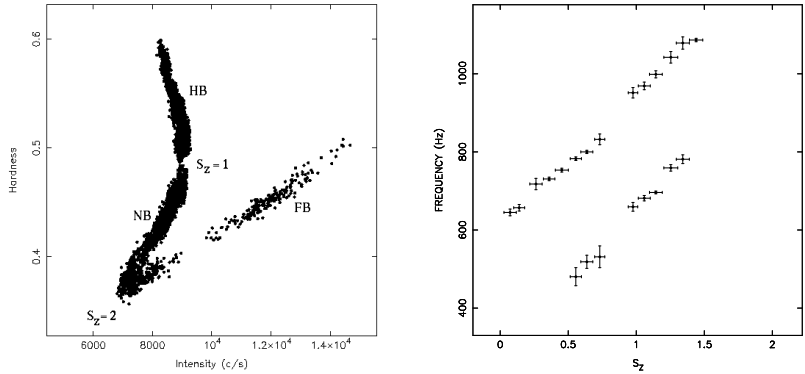

\begin{center}
\begin{tabular}{c}
\psfig{figure=gx17p2_hid_annotated.postscript,width=5cm}
\hskip0.5cm
\psfig{figure=gx17p2_freq_vs_Ix.postscript,width=5cm}
\end{tabular}
\caption{\scriptsize Z-track in the hardness-intensity diagram 
and dependence of twin peak frequencies on position $S_Z$ along the
Z-track (Wijnands et al. 1997c). The count rate starts to fall with
inferred mass accretion rate when the source enters the normal branch
at $S_Z$=1 (NB; the middle of the three branches of the Z track), but
the QPO frequencies keep on rising right through the point $S_Z$=1 all
the way up to $S_Z$=1.5, halfway down the NB. \label{17p2plots}}
\end{center}
\end{figure}

\begin{figure}[htbp]
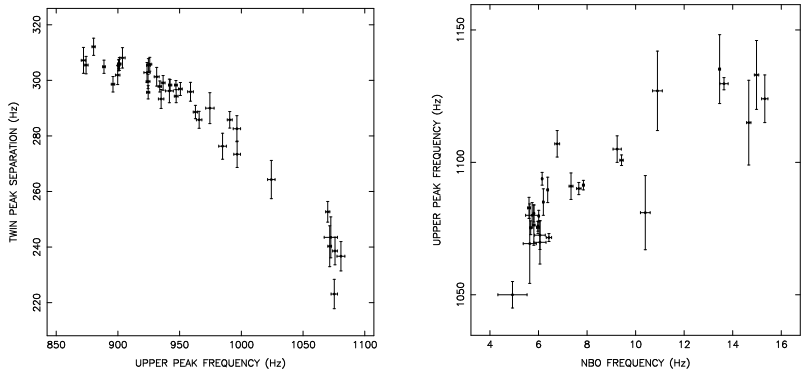

\begin{center}
\begin{tabular}{c}
\psfig{figure=scox1_separation.postscript,width=5cm}
\hskip0.5cm
\psfig{figure=scox1_khz_vs_nbo.postscript,width=5cm}
\end{tabular}
\caption{\scriptsize Sco\,X-1. Left: twin peak separation vs. the 
frequency of the upper kHz peak (Van der Klis et al. 1997b). There is
a clear decrease in the peak separation as the frequency (and inferred
mass accretion rate) increase. Right: correlation between NBO
frequency and upper kHz QPO frequency (cf. Van der Klis
et al. 1996c). \label{scoplots}}
\end{center}
\end{figure}

In most sources, the separation between the kHz peaks is consistent
with remaining constant while the peaks change frequency
(Figs.\,\ref{freq_Ix} and \ref{17p2plots}). However, in
Sco\,X-1 the peak separation varies systematically with inferred
\mdot, from \about310\,Hz when the upper peak is near 870\,Hz 
to \about230\,Hz when it is near 1075\,Hz: the peaks move closer
together by \about80\,Hz while they both move up in frequency as
\mdot\ increases (Fig.\,\ref{scoplots} {\it left}; Van der Klis et al. 1997b). 
In 4U\,1608$-$52 (M\'endez et al. 1997b) the peak separation between
two different observations also appears to have changed.

In Sco\,X-1, there is a good correlation between the frequencies of
the 6\,Hz ``normal branch oscillations'' and the kHz QPO which is not
easy to understand within current models (Fig.\,\ref{scoplots} {\it
right}; see Van der Klis et al. 1996c for a suggestion).

One of the distinguishing characteristics of kHz QPO is that they
often show a relatively large coherence. The quality factor Q, defined
as the QPO peak's centroid frequency $\nu$ divided by its full width
at half maximum $\Delta\nu$ regularly reaches values of more than 100
in one or both of the twin peaks (although Q values near 10 are also
common).  This provides a strong constraint on ``orbiting clump'' type
models (below), as lifetime broadening considerations show that the
clumps must persist over hundreds of cycles.

Because QPO amplitudes depend very strongly on photon energy
(\S\ref{energy}), great care must be taken in comparing amplitudes
between different sources or different observations, as both
differences in the effective energy band used as differences in
absorption can strongly affect the overall amplitude. Values between
$<$1 and 15\% (rms) have been reported over the full RXTE/PCA band
(2--60\,keV).

\subsection{Burst QPO}

In five atoll sources single kHz peaks have been seen during one or
more X-ray bursts, with frequencies between 363 and 589\,Hz
(Table\,1). These oscillations are not detected in each burst. Where,
in a given source, they have been seen more than once (in more than
one burst) the frequency was the same within \about1\,Hz.  In three of
these sources (4U\,1728$-$34, 4U\,1636$-$53 and KS\,1731$-$260) twin
kHz peaks in the persistent emission have also been obseved. It turns
out that in each case, the burst QPO frequencies (360--580\,Hz) are
close to the frequency {\it differences} between the twin peaks (in
4U\,1728$-$34), or twice that (in the other two sources). Of course
this immediately suggests a beat-frequency phenomenon (Strohmayer et
al. 1996c; see \S\ref{beat}). In a sixth atoll source (4U\,0614+09)
there is marginal evidence for a third peak at the twin-peak
separation frequency which corresponds to an oscillation in the
persistent emission rather than in X-ray bursts.

\begin{figure}[htbp]
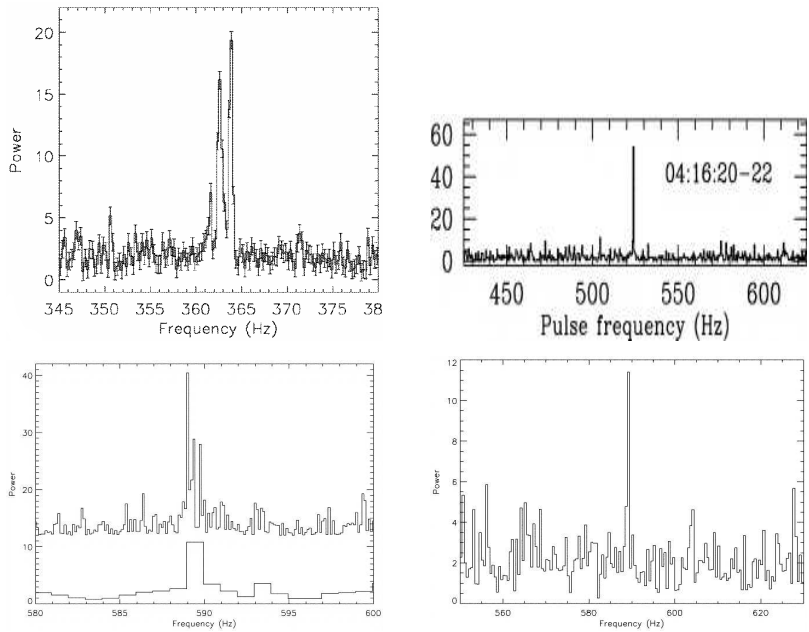

\begin{center}
\begin{tabular}{c}
\psfig{figure=apjl_xte_1728_stromayer_burstqpo.postscript,width=5cm}
\hskip0.5cm
\psfig{figure=apjl_xte_KS1731_morgan_burstqpo.postscript,width=5cm,height=3cm}
\end{tabular}
\begin{tabular}{c}
\psfig{figure=apjl_xte_1743_burstqpodrift.postscript,width=5cm}
\hskip0.5cm
\psfig{figure=apjl_xte_1743_burstqposteady.postscript,width=5cm}
\end{tabular}
\caption{\scriptsize Power spectra of burst QPO in 4U\,1728$-$34 ({\it 
top left}), KS\,1731$-$260 ({\it top right}), and the source identified
as probably MXB\,1743$-$29 ({\it bottom}) (Strohmayer et al. 1996c; 
Smith et al. 1997; Strohmayer et al. 1997). Clear drifts are visible
in the left-hand frames.\label{burstqpo}}
\end{center}
\end{figure}

The burst QPO have shown even larger coherence than the twin peaks
(Fig.\,\ref{burstqpo}). In several cases, they were coherent as long
as they could be observed (seconds). Burst QPO attained a record-level
Q of \about900 in a burst in KS\,1731$-$260 (Smith, Morgan \& Bradt
1997). In 4U\,1728$-$34 and 1743$-$29 (Strohmayer et al. 1996c, 1997),
drifts by \approxlt1\,Hz have been observed in the QPO
frequency. These drifts are suggestive of the bursting layer slightly
expanding and then recontracting, changing its rotation rate to
conserve angular momentum and thus modulating the QPO frequency.  Both
the high Q value and the small frequency drifts of the burst QPO
therefore support the view that their origin lies in the neutron star
spin, and that their frequency is close to the spin frequency. We
could be seeing hot spots at the surface related to uneven nuclear
burning caused by either a propagating nuclear burning front or the
build-up (due to magnetic channeling) of extra fuel at the magnetic
poles which burns during the burst. In the latter case the link
between the magnetic field lines and the hot spots must be severed
during the X-ray bursts in order to make the frequency drifts
possible. This could happen as a consequence of the convection in the
bursting layer (Bildsten 1996, priv. comm.).

\subsection{A beat-frequency phenomenon}\label{beat}

The three cases of three commensurate frequencies described above
demonstrate that a beat-frequency phenomenon plays a central part in
kHz QPO phenomenology: two frequencies must be interacting to produce
a third one. If the burst QPO indeed arise through surface features
spinning around with the neutron star, then the ``other'' frequency
must be due to something that is able to beat with the spin frequency
in such a way as to produce (to first order) only one additional
frequency. The simplest way to do this is by some other azimuthal
variation.  The most straightforward model is that this azimuthal
variation consists of inhomogeneities at the inner edge of the
accretion disk moving around the neutron star with the local Keplerian
angular velocity.  In this model the upper kHz peak is at the Kepler
frequency corresponding to the inner edge of the disk, and the lower
peak at its beat with the spin (assuming disk and star spin in the
same direction; otherwise the interpretation of the two kHz peaks is
reversed). The fact that the burst QPO sometimes have a frequency
twice the twin peak separation implies that in those cases we see two
events (e.g., due to two hot spots on the surface) per neutron star
spin period, where the inner edge of the disk sees only one. Other
models than this simplest one are in principle possible, for example
models involving beats between the spin and oscillations either at the
neutron star surface or in the disk.  These could have very different
angular velocities than the Keplerian one, and could have $n$-fold
symmetries making the beat frequency $n$ times higher. This kind of
devil's advocate models will need careful attention in order for
interpretations in terms of neutron star parameters and general
relativistic effects to be compelling. Even if the upper peak reflects
clump orbital motion at the inner edge of the disk, we have as yet no
independent evidence that the orbital velocity of these clumps is the
same as that of a test particle in vacuum; in fact, radiation forces
are expected in some cases to be considerable.

If we accept that some kind of beat-frequency model is at work, with
the burst QPO at the neutron star spin frequency (or twice that), the
upper kHz peak at the Kepler frequency of some preferred
orbital radius around the neutron star, and the lower kHz peak at the
difference frequency between these two, then the next question is,
what is this preferred radius.  Strohmayer et al. (1996c) suggested
that it is the magnetospheric radius, Miller, Lamb \& Psaltis (1996)
proposed it is the \nobreak sonic radius (\S\ref{beatmodel}).

\begin{figure}[htbp]
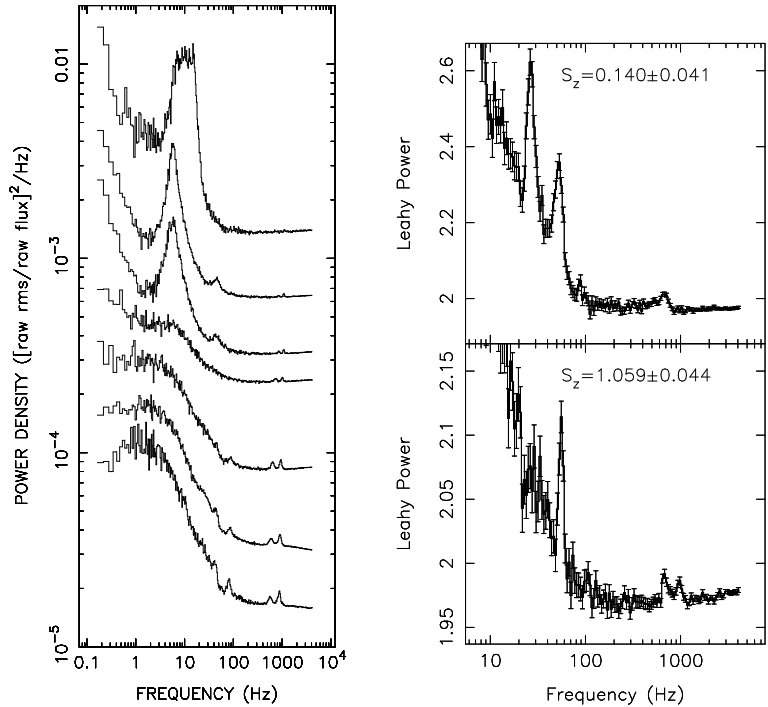

\begin{center}
\begin{tabular}{c}
\psfig{figure=apjl_xte_sco_horne_fig2.postscript,width=4.5cm}
\hskip0.5cm
\psfig{figure=apjl_xte_gx17p2_wijnands_powerspectra.postscript,width=5cm}
\end{tabular}
\caption{\scriptsize Left: simultaneous kHz QPO, HBO fundamental and
harmonic (near 45 and
90\,Hz) and NBO (near 6\,Hz) in Sco\,X-1 (Van der Klis et
al. 1997b). Right: simultaneous kHz QPO, and HBO fundamental and
harmonic in GX\,17+2 (Wijnands et al. 1997c). \label{kHzHBO}}
\end{center}
\end{figure}

In models of this kind, which involve the neutron star spin as one of
the frequencies participating in the beat-frequency process, the twin
peak separation is predicted to be constant. Therefore, the results of
a varying twin-peak separation in Sco\,X-1 (and, if confirmed,
4U\,1608$-$52) are in strong contradiction to straightforward
beat-frequency models. To maintain the beat-frequency idea, in
Sco\,X-1 a {\it second} unseen frequency has to be postulated in
addition to the (in Sco\,X-1) unseen neutron-star spin frequency. Two
different proposals to explain the systematically varying peak
separation frequency in Sco\,X-1 have been put forward.  Lamb
(priv. comm. 1996) suggests different inner disk edge frequencies at
different heights above the symmetry plane in a thick disk, White \& 
Zhang (1997) propose changing rotation frequencies of the puffed-up
neutron star envelope which maintains its angular momentum, and
therefore slows down, as the envelope puffs up, similarly to what has
been proposed to happen in X-ray bursts. Both proposals rely on the
near-Eddington accretion typical of Z sources and would not work for
4U\,1608$-$52 near 0.1\,$L_{\rm Edd}$

In all four Z sources with kHz QPO (Table\,2) twin kHz peaks and the
so-called horizontal-branch oscillations (HBO; Van der Klis et
al. 1985) are seen simultaneously (Fig.\,\ref{kHzHBO}). HBO have been
interpreted to be a product of the magnetospheric beat-frequency
mechanism (Alpar \& Shaham 1985, Lamb et al. 1985), and so have kHz
QPO in atoll sources (Strohmayer et al. 1996c). Clearly, in Z sources
this model can {\it not} explain both types of QPO. It is possible in
principle that the kHz QPO in the Z sources is a different phenomenon
from that in the atoll sources (e.g., Strohmayer et al. 1996c), but
this seems unlikely: the frequencies, their dependence on \mdot, the
coherencies, the peak separations and the fact that there are {\it
two} peaks, one of which sometimes becomes undetectable at extreme
\mdot, are too similar to attribute to just coincidence. The amplitude
of the kHz QPO in Z sources is also in the range of that seen in the
atoll sources (although in some atoll sources much higher values have
been seen). If there is only one phenomenon, then the variable twin
peak separation in Sco\,X-1 and perhaps 4U\,1608$-$52, the
simultaneous presence of kHz QPO and HBO in Z sources, {\it and} the
direct indications for a beat frequency in the atoll sources must all
be explained within the same model, a formidable challenge. If the
sonic-point beat-frequency model explains the kHz QPO and the
magnetospheric beat-frequency model the HBO, then in Z sources there
is a Keplerian disk well inside the magnetosphere (Van der Klis et
al. 1997b).

\subsection{Single peaks}\label{single}

One of the twin peaks sometimes (at high or low \mdot\ for the source
considered) drops below the detection limit while the other one is
still observable (Tables\,1 and 2). Burst QPO so far were always
single. In addition to these cases of single kHz QPO peaks, there has 
been a number of cases where the suspicion arose that a single kHz
peak might be something else than either one of the twin peaks or a
third peak located near the difference frequency of the twin peaks. In
particular this was the case in 4U\,1608$-$52 (Berger et al. 1996) and
4U\,1636$-$53 (Zhang et al. 1996, Wijnands et al. 1997a). In these
cases, the frequency of the QPO peak seemed to vary erratically over a
range of 830--890\,Hz (4U\,1608$-$52) and 835--897\,Hz (4U\,1636$-$53)
while there were no obvious variations in the X-ray colors that
indicated that \mdot\ changed.  The frequency variations seemed
unrelated to count-rate variations as well. However, in 4U\,1608$-$52
Mendez et al. (1997b) recently, by applying a new technique to
increase sensitivity to weak QPO, detected a second peak. A dependence
of frequency on count rate (although a different one during two
observations three days apart) was also found. So, although the final
word is not yet in, it now seems likely that these ``single'' peaks
\nobreak are just another aspect of the twin-peak kHz QPO phenomenon.

\subsection{Energy dependence}\label{energy}

\begin{figure}[htbp]
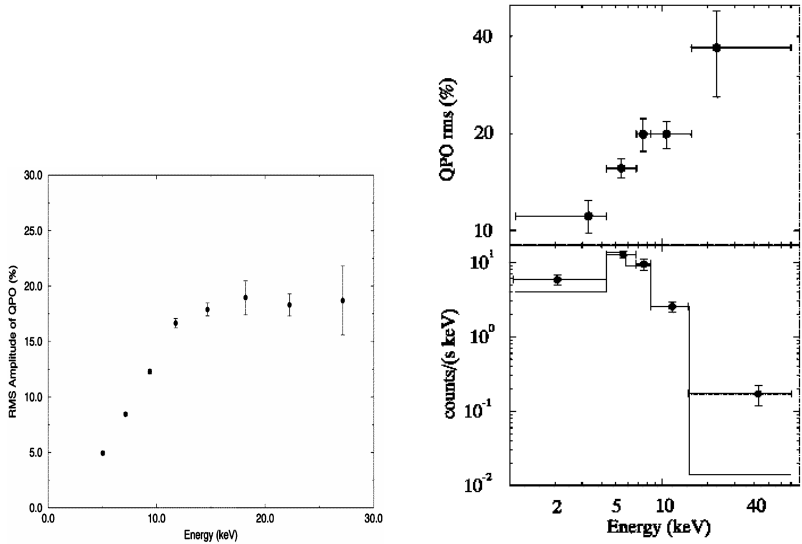

\begin{center}
\begin{tabular}{c}
\psfig{figure=apjl_xte_1608_berger_energydependence.postscript,width=5cm,height=5cm}
\hskip0.5cm
\psfig{figure=apjl_xte_0614_mendez_energydependence.postscript,width=5cm}
\end{tabular}
\caption{\scriptsize Energy dependence of kHz QPO peaks in
4U\,1608$-$52 and 4U\,0614+09 (Berger et al. 1996, M\'endez et
al. 1997a), respectively. \label{energydependence}}
\end{center}
\end{figure}

The amplitudes of kHz QPO have, in all cases where a check was
possible, shown a strong positive dependence on photon energy
(Fig.\,\ref{energydependence}). Their amplitudes when measured in a
broad photon-energy band can therefore be expected to depend strongly
on details of the low-energy part of the spectrum, which contributes
many photons and little kHz QPO amplitude: detector cutoff and
interstellar absorption will affect the overall fractional
amplitude. Reported fractional amplitudes vary between 0.5 and a few
percent in Z sources and 3 and 15\% (rms) in atoll sources when
measured over a 2--20\,keV band; at higher energies amplitudes up to
40\% (rms) have been observed.

M\'endez et al. (1997a) show that the energy spectrum of the oscillating
flux can be fitted with a blackbody spectrum with a temperature of
\about1.6\,keV and a radius of 0.5$\pm$0.2\,km 
(Fig.\,\ref{energydependence}). The QPO could therefore be due to
modulation of emission from a region on the neutron star surface with
these properties. However, the data allow many different spectral
models. For example, alternative interpretations that fit the data as
well are that the oscillations are caused by variations with an
amplitude of 2.5\% in the temperature of a \about1.1\,keV blackbody
with a radius of \about10km, or by \about5\% variations in the optical
depth of an unsaturated Comptonization spectrum.

\begin{figure}[htbp]
\begin{center}
\begin{tabular}{c}
\psfig{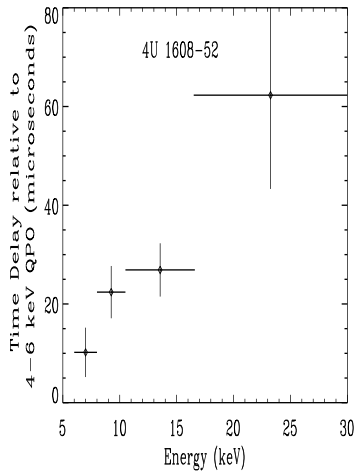}
\end{tabular}
\caption{\scriptsize Time lags as a function of photon energy in the
kHz QPO of 4U\,1608$-$52 (Vaughan et al. 1997). \label{lags}}
\end{center}
\end{figure}

A final strong model constraint is provided by the small magnitude of
{\it time lags} between the kHz QPO signal as observed in different
energy bands. Time-lag measurements require very high signal-to-noise
ratios, and have so far only been made in the ``single'' peaks in
4U\,1608$-$52 and 4U\,1636$-$53 near 850\,Hz, and in a 730\,Hz peak in
4U\,0614+09 which was probably an upper peak (Vaughan et
al. 1997). Finite lags of 10--60$\mu$sec were discovered in
4U\,1608$-$52; the hard photons lag the soft ones by increasing
amounts as the photon energy increases (Fig.\,\ref{lags}). Upper
limits of 30$\mu$sec and 45$\mu$sec were set in 4U\,1636$-$53 and
4U\,0614+09, respectively. These are by far the smallest lags ever
measured; they correspond to light travel distances of 3--20\,km! For
rather general assumptions about the spectral formation mechanism,
this limits the scale of any Compton scattering regions dominating the
spectral shape to between a few and a few tens of km.


\subsection{Puzzles}\label{puzzles}

The great enigma in the phenomenology right now is, in my opinion, the
peculiar lack of correlation between kHz QPO frequency and average
source luminosity, whereas {\it in each individual source} a strong
correlation between frequency and \mdot\ is observed (Van der Klis
1997). In 4U\,0614+09, at a luminosity of a few times
10$^{-3}\,L_{\rm Edd}$, similar QPO frequencies have been observed as in
4U\,1820$-$30, which is near 10$^{-1}\,L_{\rm Edd}$, and in Sco\,X-1,
which is inferred to be a near-Eddington accretor. (It is interesting
to note that no kHz QPO have been detected in sources such as GX\,9+1,
GX\,9+9 and GX\,13+1 which are thought to cover the
0.1--0.5\,$L_{\rm Edd}$ range.) This must mean that another, compensating,
parameter than just instantaneous accretion rate is affecting the
properties of the kHz QPO, most likely by directly affecting the
frequency, although some kind of selection effect that leads to
suppression of any QPO outside the 300--1200\,Hz range is also a
possibility. This latter possibility of course requires that the peaks
actually observed in sources with different luminosities are in some
sense ``different''. One would expect that in sources that go through
a large decrease in accretion rate (transients) several ``new'' QPO
peaks would successively appear near 1200\,Hz, move down in frequency
and disappear near 300\,Hz. This has not been seen and seems somewhat
unlikely.

An obvious candidate for such a compensating parameter is the neutron
star magnetic field strength, but neutron star mass or spin, either by
their effects on the surrounding space-time or directly, might play a
role as well. What would be required, specifically, is that there
exists a correlation or an anti-correlation between, say, the magnetic
field strength $B$ of the neutron star and its mean accretion rate
$\langle\dot M\rangle$, and that the QPO frequency depends on $B$ in
such a way as to approximately compensate the \mdot\
effect. Interestingly, it has been concluded previously (Hasinger and
Van der Klis 1989, see Van der Klis 1995) on the basis of comparing Z
and atoll source phenomenology that $\langle\dot M\rangle$ and $B$ are
correlated among LMXBs. Recently, spectral modeling (Psaltis et
al. 1997) has tended to confirm this, and very recently White \& 
Zhang (1997) have shown that if one assumes a standard spin-up law
(which, as they remark, may be incorrect, see below) and spin up at an
$\langle\dot M\rangle$ equal to the current one, the spin rates that
can be inferred from the kHz QPO imply a tight correlation between
average $\langle\dot M\rangle$ and $B$. In the magnetospheric
beat-frequency model (Alpar \& Shaham 1985) $B$, if it really is
correlated with $\langle\dot M\rangle$, qualitatively fits the
requirements for the compensating parameter sketched above. However,
the simultaneous detections of HBO and kHz QPO in the Z sources
(\S\ref{beat}) as well as a number of other considerations (Miller et
al. 1996) make this model unattractive. Perhaps, the magnetic field
strength affects the inner accretion flows in other ways than by just
terminating the disk at the magnetospheric radius. If magnetic
stresses could somehow slow down the (for example, orbital) motion
responsible for the kHz QPO, that would do the trick. Of course,
radiative stresses diminish the effective gravity and are expected to
slow down orbital motion (Miller \& Lamb 1993), but the luminosity is
not independent from \mdot, but instead expected to vary
proportionally to it, so that radiative stresses cannot fulfill the
compensating-parameter role: we already know that when in a given
source
\mdot\ goes up so does $L_{\rm X}$, but this does not prevent the QPO
frequency from going up as well.

Another suggestion (A. King, at the September 1997 Amsterdam Workshop
on Compact Stars and Close Binary Systems) is that the neutron star
surface temperature (as set by some longer-term average over \mdot)
affects the frequency-\mdot\ relation. The attraction of this idea is
that it might also explain the {\it changes} in the frequency
vs. count-rate relation in 4U\,0614+09 and 4U\,1608$-$52
(\S\ref{twinpeaks}).

Two final puzzles: (i) The maximum kHz frequencies observed in each
source (Tables\,1 and 2) are nearly all constrained to the relatively
narrow range 1036--1228\,Hz. In at least five sources, spread over the
entire range of average X-ray luminosity, the upper peak has been
observed to disappear below the detection limit when its
frequency is somewhere in this range as soon as the flux exceeds a
certain limit. However, this flux limit is widely different between
sources. There is no evidence for a sudden drop in QPO amplitude near
these flux limits. It is as if this clustering of frequencies is
caused by a conspiracy for the apparently brighter sources to have
intrinsically weaker kHz QPO, which can't be true. (ii) The 12 neutron
star spin rates that can be inferred from the kHz QPO (Tables\,1 and
2; spin is estimated as either the burst QPO frequency or half that,
or the twin peak separation frequency) are also confined to a quite
narrow range (235--363\,Hz). According to White \& Zhang (1997) the
least contrived explanation for this is that the magnetospheric radius
(and therefore the equilibrium spin rate) depends only very weakly on
accretion rate, and the magnetic field strengths of these neutron
stars are all similar.

\subsection{Promises}\label{promises}

From the above it should be clear that kHz QPO likely provide the
first measurements of the spin rates of low-magnetic-field neutron
stars in LMXB --- a holy grail of X-ray astronomy since twenty years.
The spin rates are in rough accordance with expectations from the
LMXB-millisecond radio pulsar evolutionary connection. As was already
discussed in \S\ref{puzzles}, they may also present a possibility to
constrain neutron star $B$ fields.

\begin{figure}[htbp]
\begin{center}
\begin{tabular}{c}
\psfig{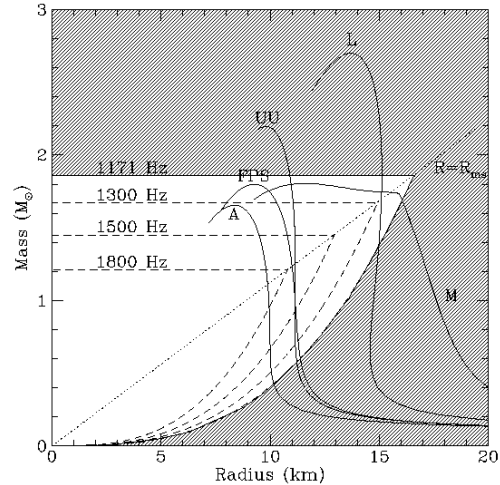}
\end{tabular}
\caption{\scriptsize Allowed region in the mass-radius diagram if a
Keplerian frequency of 1171\,Hz exists. The hatched area is
excluded. After Miller et al. 1996. \label{wedge}}
\end{center}
\end{figure}

The most exciting aspect of kHz QPO, however, is that potentially they
can be used to constrain neutron-star masses and radii, and to test
general relativity. There is a lively ongoing discussion about
this. Kaaret, Ford \& Chen (1997) have proposed that the behavior of
the ``single'' QPO peaks in 4U\,1608$-$52 and 4U\,1636$-$53 described
above is related to orbital motion near the marginally stable orbit,
and from this derive neutron star masses of \about2\msun. Zhang,
Strohmayer \& Swank (1997) have proposed that the narrow range of
maximal frequencies described above must be identified with the
general-relativistic marginally stable frequencies, which leads them
to the conclusion that the neutron stars' masses are near 2\msun\ as
well. An alternative possibility is that the maximal frequencies are
set by the Keplerian frequency at the neutron star surface. This
requires the star to be larger than the marginally stable orbit and
for
\about1.4\msun\ neutron stars would favour the stiffest equations of
state. As should be clear from \S\ref{single} and \ref{puzzles}, there
are quite considerable problems with these interpretations. Very
recently, Stella \& Vietri (1997) have proposed that the
Lense-Thirring precession frequencies for the inner edge of the disk
that can be predicted on the basis of the measured neutron-star spin
frequencies and Kepler frequencies can be seen in the power spectra of
some atoll and perhaps even Z sources in the form of broad low-frequency
humps (these humps were already noted by Hasinger \& Van der Klis
1989 and are very clear in some RXTE data, e.g., Fig.\,\ref{discovery}
{\it right}).

Just the assumption that the upper peak corresponds to Keplerian
motion around the neutron star allows to set useful limits on neutron
star parameters, a point made by Miller et al. (1996) in their paper
on a particular model that interprets the upper peak in this way (see
\S\ref{beatmodel}). Different from the proposals just mentioned, these limits do
{\it not} rely on identifying any of the observed frequencies with the
marginally stable orbital frequency. Kluzniak \& Wagoner (1985) and
Kluzniak, Michelson \& Wagoner (1990) already anticipated that if the
orbital frequency of accreting matter near a neutron star could be
measured, this could be used to to determine neutron star parameters
and test general relativity, although these authors did not present
the detailed argument below. There are two direct constraints on the
neutron star mass and radius from the simple assertion that there is
stable Keplerian motion at the frequency $\nu_{\rm u}$ of the upper peak:
(1) the radius of the star $R$ must be smaller than the radius of this
Keplerian orbit, in Schwarzschild coordinates $R <
(GM/4\pi^2\nu_{\rm u}^2)^{1/3}$, and (2) the radius of the marginally stable
orbit must {\it also} be smaller than this: $6GM/c^2 <
(GM/4\pi^2\nu_{\rm u}^2)^{1/3}$ or $M < c^3/(2\pi6^{3/2}G\nu_{\rm u})$, as no
stable orbit is possible within this radius.  Condition (1) is a
mass-dependent upper limit on the radius of the star, and condition
(2) provides an upper limit on the mass. \hide{In a Schwarzschild
geometry these expressions are exact TRUE?} Fig.\,\ref{wedge} shows
these limits in the neutron star mass-radius diagram for
$\nu_{\rm u}=1171$\,Hz, plus an indication of how the excluded area ({\it
shaded}) would shrink for higher values of the upper peak frequency
(the currently highest value is 1228\,Hz; Table\,1). If the softest
EOS are incorrect, it seems likely that this method will eventually
demonstrate this. Putting in the corrections for the frame dragging
due to the neutron star spin (which were neglected in
Fig.\,\ref{wedge}) requires knowledge of the spin rate (presumably the
twin peak separation, or half that). The correction also depends
somewhat on the neutron star model, which determines the relation
between spin rate and angular momentum, so that the limits as plotted
in Fig.\,\ref{wedge} become slightly different for each EOS. These
corrections change the limits on mass and radius only slightly. For a
Keplerian frequency of 1193\,Hz (Wijands et al. 1997a) the above
equations imply $M<1.84$\msun\ and $R_{\rm NS}<16.3$\,km. With frame
dragging corrections for an assumed spin rate of 275\,Hz these limits
change into $M<2.1$\msun\ and $R_{\rm NS}<16.5$\,km for a wide range of
equations of state (Wijnands et al. 1997a).

\section{Models}\label{models}

This section is intended not so much to provide an exhaustive
comparative discussion of the models that have so far been proposed
for kHz QPO, as to alert the reader to the various basic physical
pictures that have been discussed.

Of course, the phenomenology as described in the previous section very
strongly suggests that a beat-frequency model of some kind is at
work. Neutron star spin and disk Keplerian motion are periodic
phenomena known to be present in the system and are therefore natural
candidates for providing the basic frequencies. However, it is too
early to declare any proposed implementation of a beat-frequency model
for kHz QPO an unqualified success. Let us first look at other models
that have been put forward.

\subsection{Neutron star vibrations}

Remarkably short shrift has been given so far to neutron star
vibration models. The short time scale variations in kHz QPO frequency
and the lack of higher-frequency peaks have been cited as reasons for
rejecting these models. Of course, a fundamental problem for vibration
models (also for lower-frequency QPO) has always been to explain how
the vibrations would be able to produce appreciable modulations of the
X-ray flux, in the case of kHz QPO with amplitudes up to 15\% (full
band) to 40\% ($>$10\,keV). Also, it is unclear what physics would be
required to pick out just 2 (or 3) frequencies from the range of modes
one would expect in most of these models.

\subsection{Photon bubble model}

A model based on numerical radiation hydrodynamics has been proposed
by Klein et al. (1996) for the case of the kHz QPO in Sco\,X-1. In
this model accretion takes place by way of a magnetic accretion
funnel. In the funnel, the mass accretion rate is locally
super-Eddington, and photon bubbles form, which rise up by buoyancy
through the accreting matter. 

\begin{figure}[htbp]
\begin{center}
\begin{tabular}{c}
\psfig{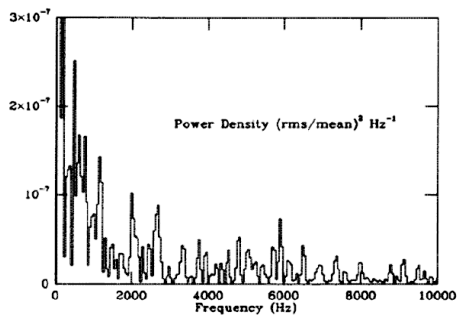}
\end{tabular}
\caption{\scriptsize Theoretical photon bubble oscillation power
spectrum (Klein et al. 1996). \label{pbo}}
\end{center}
\end{figure}

Klein et al. find that the ``bursting of
the bubbles'' at the top of the flow can produce kHz QPO signals that
resemble the observed ones in several respects
(Fig.\,\ref{pbo}). Their initial calculations, for a total X-ray
luminosity of 0.17$L_{\rm Edd}$ predicted that there should be QPO peaks
at higher frequencies. High time-resolution observations of Sco\,X-1
performed in order to check on this showed no evidence of
higher-frequency peaks down to quite good limits (Van der Klis et
al. 1997b; this provides a strong constraint on various other models
as well). In the most recent calculations (for $B$=5\,10$^9$\,Gauss and
0.24$<L/L_{\rm Edd}<$0.4; Klein 1997, private communication) one or two
strong QPO peaks are found, whose frequencies increase with accretion
rate, and the higher-frequency peaks are of low amplitude. The model
also predicts the presence of a power law at frequencies $>$2\,kHz
with an index of \about5/3. These models are not likely to be able to
naturally produce beat frequencies such as observed in atoll
sources. Perhaps they apply to the Z sources. This would require the
atoll and Z source kHz QPO to have different physical origins.

\subsection{Jets}

The relation between the QPO frequencies observed in Sco X-1 can be
nicely explained with a model where each of the two QPO signals comes
from one of two diametrically opposed relativistic jets emanating from
the central source. In this picture a central shrouded X-ray pulsar
provides the basic high-frequency signal. The observer sees the pulsar
signal reflected off inhomogeneities in the two jets. Three frequency
shifts affect the observed frequencies: a redshift (identical for each
jet) because the inhomogeneities move away from the pulsar at
relativistic speed, special-relativistic time dilation, and an
additional redshift and blueshift for the receding and approaching
jet, respectively. The predicted frequencies for the signals reflected
off the two jets will be
$\nu_\mp=\nu_{\rm pulse}(1-v/c)/(1\pm(v/c)\cos\theta)$, where $v$ is the
jets' speed, $\theta$ their angle with the line of sight, and
$\nu_{\rm pulse}$ the unseen pulse frequency (Van der Klis et al. 1997b).
This model fits the Sco\,X-1 data remarkably well, for
$\nu_{pulse}$=1370\,Hz (which could be twice the neutron star spin
frequency) and $\theta$=61$^\circ$, if we assume that with increasing
\mdot, the jet's speed decreases from $v/c$=0.48 to 0.26 (Van der Klis
et al. 1997b). However, it is hard to see how this model can explain
the atoll sources' kHz QPO properties, as frequency shifts {\it
without} the peaks' frequency separation being affected, as are
observed in those sources, can not be accommodated by the model.

\subsection{Beat-frequency models}\label{beatmodel}

Finally, let's turn to beat-frequency models. Although numerous disk
and star frequencies could be hypothesized to beat together, the two
beat-frequency models that have been discussed both identify the upper
peak's frequency with the Keplerian frequency of the accretion disk at
some preferred radius, and the lower peak with the beat between this
Keplerian frequency and the neutron star spin frequency. When disk and
star spin in the same direction, this naturally leads, to first order,
to the generation of only one beat frequency. The magnetospheric
beat-frequency model, which has been used previously to explain
certain cataclysmic variable oscillations (Patterson 1979) and the
so-called horizontal-branch oscillations (HBO) in Z sources (Alpar \& 
Shaham 1985; Lamb et al. 1985), uses the magnetospheric radius $r_{\rm M}$
as the preferred radius of which we observe the Kepler frequency. As
HBO and kHz QPO have been seen {\it simultaneously} in all four Z
sources where kHz QPO have so far been observed, at least {\it one}
additional model is required.

According to Miller, Lamb \& Psaltis (1996), applying the
magnetospheric beat-frequency model to the kHz QPO leads to several
difficulties. They propose the {\it sonic-point model} instead. This
model uses the sonic radius as the preferred radius of which we
observe the Kepler frequency. The sonic radius is defined as the
radius where the radial inflow velocity becomes supersonic. In
the absence of other stresses, the sonic radius is located near the
general-relativistic marginally stable orbit (at $6GM/c^2$ in a
Schwarzschild geometry and closer in in a Kerr geometry). Radiative
stresses may change the location of the sonic radius, as indeed is
required by the observation that the kHz QPO frequencies vary with
\mdot. As we have seen (\S\ref{promises}), interpretations along 
these lines have direct consequences for the EOS of high-density
matter and provide possibilities to test general relativity in the
strong-field regime.

\begin{figure}[htbp]
\begin{center}
\begin{tabular}{c}
\psfig{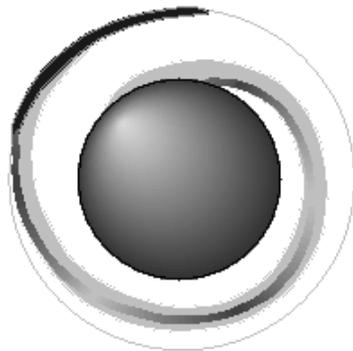}
\end{tabular}
\caption{\scriptsize The spiral flow in the Miller et al. (1996)
model. \label{spiralflow}}
\end{center}
\end{figure}

In the sonic-point model, the mechanism that produces the upper peak
is as follows. At the sonic radius clumps form in the accretion flow
that live for tens to hundreds of QPO cycles (i.e., several 0.01 to
several 0.1\,s). During a clump's lifetime, its matter gradually
accretes onto the neutron star surface. This accreting matter follows
a fixed spiral-shaped trajectory in the frame corotating with the
Keplerian motion, and therefore hits the neutron star surface at a
specific spot, the ``footpoint''of that clump's spiral flow
(Fig.\,\ref{spiralflow}). Enhanced accretion at the footpoint produces
enhanced emission there, and because the footpoint travels around the
neutron star surface at the Keplerian angular velocity of the sonic
radius, irrespective of what the neutron star's spin is, the observer
sees the hot spot change aspect and perhaps appear and disappear with
the sonic radius' Keplerian frequency. The narrowness of the QPO peaks
implies that all the clumps are near one precise radius, allows for
relatively little fluctuations in the spiral flow, and requires the
above-mentioned long clump lifetimes.

In Z sources, applying this model for the kHz QPO, and the
magnetospheric beat-frequency model for the HBO leads to the
conclusion that the sonic radius $r_{\rm S}$ is well within the
magnetosphere (Van der Klis et al. 1997b), so that only part of the
matter can be ``picked up'' by the magnetic field lines at $r_{\rm M}$, and
the remainder must maintain its Keplerian flow to much closer in. The
modulation of the radiation caused by this mechanism is not a
modulation of the total luminosity, but a modulation of the direction
into which this luminosity is emitted (``beaming''). If some of these
neutron stars have a magnetic field strong enough to channel some of
the accreting matter to the magnetic poles (as required by the
magnetospheric beat-frequency model for Z sources), then {\it another}
beaming modulation is expected at the neutron star spin
frequency. However, so far {\it this} modulation has not been detected
in any Z source, which presents the difficulty that one has to somehow
get rid of the spin frequency but not of the sonic-radius Keplerian
frequency. As both are beaming modulations and the frequencies are
similar, this may require some finetuning of the scattering process
that is smearing the pulsations.

Miller et al. (1996) predict that as the sonic radius approaches the
general-relativistic marginally stable orbit the frequency of the
upper peak will hit a ``ceiling'' and remain stable for further
increases in accretion rate. There are so far no data that have shown
this. Instead it has been observed that the QPO gradually weaken as
\mdot\ increases, and finally disappear below the detection limit above some
level of inferred accretion rate. This \mdot\ level, however, is very
different between sources (much higher in sources with a higher
average luminosity), but at frequencies that are mostly in the range
1100--1200\,Hz.  Perhaps this is what {\it really} happens when
$R_{\rm MSO}$ is reached.

Now let's turn to the question how the lower peak is produced in the
sonic-point model. The footpoint running over the neutron star surface
will of course encounter the same point on the surface once every beat
period between the neutron star spin and the sonic-point Keplerian
frequencies. If there are $B$ gradients over the surface this could
affect the emission properties of the footpoints at the
beat frequency. However, this is {\it not} the model proposed by
Miller et al. They propose instead that the physical interaction that
eventually leads to the modulation of the X-ray flux at the
beat frequency takes place at the sonic radius. In their model, X-rays
produced by the accretion of matter channeled onto the magnetic poles
are emitted into two broad ``lighthouse beams'', sweeping around at
the neutron star spin frequency. These beams irradiate the clumps at
the sonic radius when they sweep over them, which happens once per
beat period. This leads to a modulation, at the beat frequency, in the
rate at which the clumps provide matter to their spiral
flows. Consequently, the accretion of matter onto the footpoints, and
therefore their emission, is modulated at the beat frequency, and this
leads to the lower peak we see in the power spectra. This model
predicts various aliases and harmonic of the observed peaks should
also be present, but these have not been observed so far. The model
requires the pulsar beams to extend out to the sonic radius with
sufficient strength to affect the accretion flow there, yet to be
smeared, presumably by scattering further out, to below the detection
levels that have so far been reached (e.g., Vaughan et al. 1997). On
the other hand the footpoints' emission beams must be able to
propagate to infinity in order for us to see the sometimes quite
strong upper peaks. As double peaks are seen in sources between
several 10$^{-3}$ and \about1\,$L_{\rm Edd}$, all these processes must
operate in a way to keep the phenomenology the same over a large range
in \mdot\ (but this, of course, is a problem in any interpretation
that attempts to explain these phenomena by one mechanism across the
board). The cases of 4U\,1636$-$53 and KS\,1731$-$260, where the
pulsar frequency appears to be twice the twin-peak separation
frequency, in the Miller et al. model requires a further explanation
of the question why pulsar frequency observed during X-ray bursts is
{\it twice} the interaction frequency with the sonic-radius
clumps. If, for example, we see nuclear fuel accumulated at two
magnetic poles burning in the bursts, why doesn't emission caused by
the accretion of this fuel onto both these poles interact with the
clumps, but only that from one pole?

Obviously, a large amount of effort is still required to make any of
the models so far proposed stick. Fortunately, as it looks now the
theoretical efforts that are underway at this point will be guided by
a very constraining body of RXTE data. Eventually, most LMXBs will
likely exhibit the new phenomenon, and many of its properties can be
measured with RXTE with great precision.

\acknowledgements
This work was supported in part by the Netherlands Organization for
Scientific Research (NWO) under grant PGS 78-277 and by the
Netherlands Foundation for Research in Astronomy (ASTRON) under grant
781-76-017. 

\def\lw{Lewin, W.H.G.}
\def\vpj{Van Paradijs, J.}
\def\mk{Van der Klis, M.}
\def\aj{{AJ}}                   
\def\araa{{ARA\&A\ }}             
\def\apj{{ApJ\ }}                 
\def\apjl{{ApJ\ }}                
\def\apjs{{ApJS\ }}               
\def\ao{{Appl.~Opt.}}           
\def\apss{{Ap\&SS}}             
\def\aap{{A\&A\ }}                
\def\aapr{{A\&A~Rev.}}          
\def\aaps{{A\&AS}}              
\def\azh{{AZh}}                 
\def\baas{{BAAS}}               
\def\jrasc{{JRASC}}             
\def\memras{{MmRAS}}            
\def\mnras{{MNRAS}}             
\def\pra{{Phys.~Rev.~A}}        
\def\prb{{Phys.~Rev.~B}}        
\def\prc{{Phys.~Rev.~C}}        
\def\prd{{Phys.~Rev.~D}}        
\def\pre{{Phys.~Rev.~E}}        
\def\prl{{Phys.~Rev.~Lett.}}    
\def\pasp{{PASP}}               
\def\pasj{{PASJ}}               
\def\qjras{{QJRAS}}             
\def\skytel{{S\&T}}             
\def\solphys{{Sol.~Phys.}}      
\def\sovast{{Soviet~Ast.}}      
\def\ssr{{Space~Sci.~Rev.}}     
\def\zap{{ZAp}}                 
\def\nat{{Nat\ }}              
\def\iaucirc{{IAU~Circ.}}       
\def\aplett{{Astrophys.~Lett.}} 
\def\apspr{{Astrophys.~Space~Phys.~Res.}}
\def\bain{{Bull.~Astron.~Inst.~Netherlands}} 
\def\fcp{{Fund.~Cosmic~Phys.}}  
\def\gca{{Geochim.~Cosmochim.~Acta}}   
\def\grl{{Geophys.~Res.~Lett.}} 
\def\jcp{{J.~Chem.~Phys.}}      
\def\jgr{{J.~Geophys.~Res.}}    
\def\jqsrt{{J.~Quant.~Spec.~Radiat.~Transf.}}
\def\memsai{{Mem.~Soc.~Astron.~Italiana}}
\def\nphysa{{Nucl.~Phys.~A}}   
\def\physrep{{Phys.~Rep.}}   
\def\physscr{{Phys.~Scr}}   
\def\planss{{Planet.~Space~Sci.}}   
\def\procspie{{Proc.~SPIE}}   

\end{document}